\documentclass[journal]{IEEEtran}
\ifCLASSINFOpdf
\else
\fi
\usepackage{cite}
\usepackage{url} 
\usepackage{graphicx}
\usepackage{pifont}
\usepackage{booktabs}
\usepackage{multirow}
\usepackage{amssymb}
\usepackage{rotating}
\usepackage{tabularx}
\usepackage{enumerate} 
\usepackage{breakurl}

\begin{document}
\title{Internet of Things: Survey on Security and Privacy}
\author{
        Diego Mendez\textsuperscript{1},~\IEEEmembership{}%Graduate Student Member}
        %Sahithya Kodam, 
        %Faheem Zafari, 
        Ioannis Papapanagiotou\textsuperscript{2},~\IEEEmembership{}%Senior Member,~IEEE} 	
        Baijian Yang\textsuperscript{3}~\IEEEmembership{}%Member,~IEEE}
        \\\textit{Purdue University}
        
        \thanks{1. Diego Mendez is with Purdue University, West Lafayette, IN, 47907. \protect\\
E-mail: \texttt{dmendezm@purdue.edu}}
\thanks{2. I. Papapanagiotou is with Netflix Inc., Los Gatos, CA 95032 and Purdue University, West Lafayette, IN, 47907.\protect\\
E-mail: \texttt{ipapapa@ncsu.edu}}
        \thanks{3. Baijian Yang is with Purdue University, West Lafayette, IN, 47907. \protect\\
        E-mail: \texttt{byang@purdue.edu}}
        } % <-this % stops a space

\markboth{IoT Security}%
{Shell \MakeLowercase{\textit{et al.}}: Bare Demo of IEEEtran.cls for Journals}
\maketitle
\begin{abstract} %added by Faheem Zafari, Sahithya Kodam, Diego Mendez
The Internet of Things (IoT) is intended for ubiquitous connectivity among different entities or ``things''. While its purpose is to provide effective and efficient solutions, security of the devices and network is a challenging issue. The number of devices connected along with the ad-hoc nature of the system further exacerbates the situation. Therefore, security and privacy has emerged as a significant challenge for the IoT. In this paper, we aim to provide a thorough survey related to the privacy and security challenges of the IoT. This document addresses these challenges from the perspective of technologies and architecture used. This work focuses also in IoT intrinsic vulnerabilities as well as the security challenges of various  layers based on the security principles of data confidentiality, integrity and availability. This survey analyzes articles published for the IoT at the time and relates it to the security conjuncture of the field and its projection to the future.  
\end{abstract}
\IEEEpeerreviewmaketitle
\section{Introduction} %added by Sahithya Kodam, edited by Diego Mendez
\IEEEPARstart{T}{he} essence of the Internet of Things (IoT) is the concept of every device blending with the existence of human beings. It is the state wherein there is no distinguishable difference between the operation of devices surrounding us and our actions. This means that devices become part of our experience. There is a seamless integration between us and the ``things'' around us. The various devices communicate intelligently with one another to execute daily operations. There is minimal human intervention for the operation of devices. Every device is connected to every other device, communication with one another, transferring data,  retrieving data and intelligently responding, triggering actions. The successful implementation of the IoT involves consideration of a huge number of aspects. These involve the technology used for communication, various communication protocols which form the backbone of the IoT, standards to be used for communication, hardware and embedded devices used to build the hardware, the software, operating system that is compatible with hardware and the protocols being used. By the end of 2020 it is said that there would be around 20 billion connected devices \cite{gartner2016prediction}. The data exchanged over the network will be greater than 40 Zettabytes for the same period \cite{forbes2016prediction}.  
This brings up an important discussion regarding all the data generated, stored or transmitted by IoT devices, its security and how this relates to the privacy of the users. Every approach of IoT system must be secure and provide the necessary controls and privacy to the users. Successful implementation of an IoT system is possible only when the systems are built with security as one of the central aspects of the IoT. This paper discusses the security issues at different levels of the IoT system. It presents an exhaustive survey of security and privacy issues existing in IoT systems, its enabling technologies and protocols. This paper elaborates on the current status of the field, providing the big picture based on IoT Architecture, analyzes the security challenges and vulnerabilities of the different underlying technologies and protocols, discusses the IoT security concerns under the security triad perspective and explores the current privacy issues of IoT systems under different points of view. Finally, summarizes the content to give a clear perception of the ongoing security challenges of the IoT and overlays some solutions.  
\hfill 
\hfill 

\section{Structure of IoT Systems} %Added by Diego Mendez
The IoT heterogeneous essence, dynamics, intelligence, mobility and undefined perimeters makes it a high-demand technology domain but also makes the IoT vulnerable and risky under security terms. The different platforms where the IoT is available makes it even more difficult for security researchers to find comprehensive solutions to the current security challenges. Therefore, the importance of understanding the  foundation and the components of the IoT becomes paramount.

%figure arq
\begin{figure*}[!ht]
\centering 
\includegraphics[width=13cm]{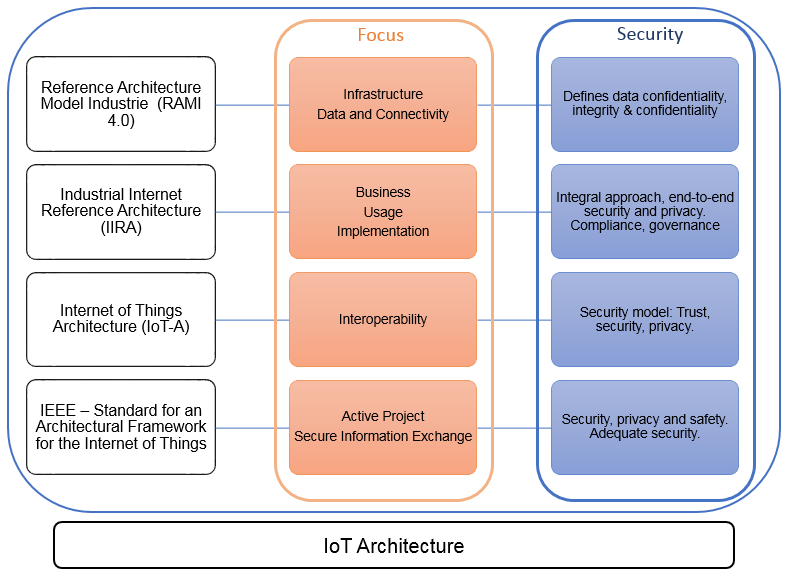}
\caption{\label{fig:IoTArqGraph}Current IoT Architecture Frameworks/Standards}
\end{figure*}

The foundation for ubiquitous computing, whose goal is to connect everyday life objects to the network using technological platforms, is made up of three components \cite{gubbi2013internet}: 
\begin{enumerate}[(a)]
\item Hardware
\item Middleware
\item Presentation
\end{enumerate}
Also, the same pattern can be observed when determining the paradigms of the IoT, according to \cite{atzori2010internet} and \cite{gubbi2013internet} three factors can be attributed to the IoT environment, those are: 
\begin{enumerate}[(a)]
\item Internet-oriented
\item Things-oriented
\item Semantic-oriented
\end{enumerate} 
Therefore, the same concept can be applied to the IoT structure. The IoT architecture, according to \cite{zhao2013survey} is composed of three layers: 
\begin{enumerate}[(a)]
\item The perception layer
\item The network layer
\item The application layer
\end{enumerate} 
The perception layer gathers environmental data, the network layer, which is composed of wired and wireless systems, processes and transmits the input obtained by the perception layer supported by technological platforms. The application layer consists of abstracted solutions that interact with the final users in order to satisfy their needs. The IoT requires architectural solutions that can manage heterogeneous states in order to work efficiently and effectively \cite{weyrich2016reference}. Figure \ref{fig:IoTArqGraph} summarizes current IoT architecture frameworks/standards and highlights their security objectives.

However, there is no unified view of the IoT framework. Some engineering bodies, including the IEEE and ETSI, have issued technology-specific standards including security guidelines \cite{zhao2013survey}. These standardization efforts have also brought up other initiatives for unified architecture and modeling, for instance the Reference Architecture Model Industrie 4.0 (RAMI 4.0) \cite{adolphs2015reference}, the Industrial Internet Reference Architecture (IIRA) \cite{industrial2015industrial} and the Internet of Things - Architecture (IoT-A) \cite{heu2013internet}. Architecture and model implementation helps IoT developers to focus and structure their efforts on users' requirements, which include connectivity, device management, data collection and analysis, scalability and security. Nevertheless, additional unification attempts are needed for simplification, always taking security communications as the main actor and enabler of IoT initiatives \cite{weyrich2016reference}. Besides the industrial domain, the scientific community has been a main contributor of the standardization of IoT protocols and technology as well \cite{atzori2010internet}. The author of \cite{weber2015internet} advocates for the need of a security-based architecture, which is lacking at the moment, where resiliency, authentication access restriction and privacy are important requirements for the future. Also, the authors of \cite{khan2012future} back a reliable architecture that address security and service requests. From a different perspective, the authors of \cite{ning2011future} promote the importance of robust and reliable standards to conduct shielded IoT architectures, currently required with insistence from the security community.  

%Figure landscape
\begin{figure*}[!ht]
\centering 
\includegraphics[width=18cm]{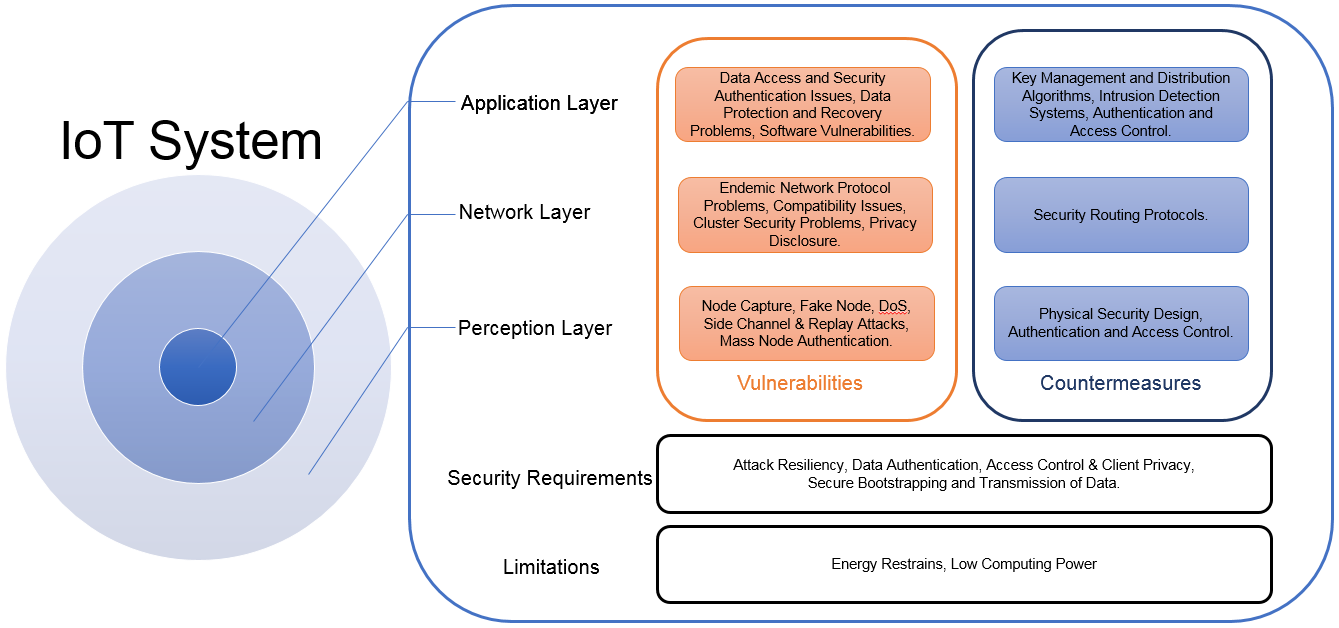}
\caption{\label{fig:IoTSecGraph}Internet of Things Security Landscape.}
\end{figure*}

\section{Vulnerable Landscape} %Added by Diego Mendez

\begin{figure*}[!ht]
\centering 
\includegraphics[width=18cm]{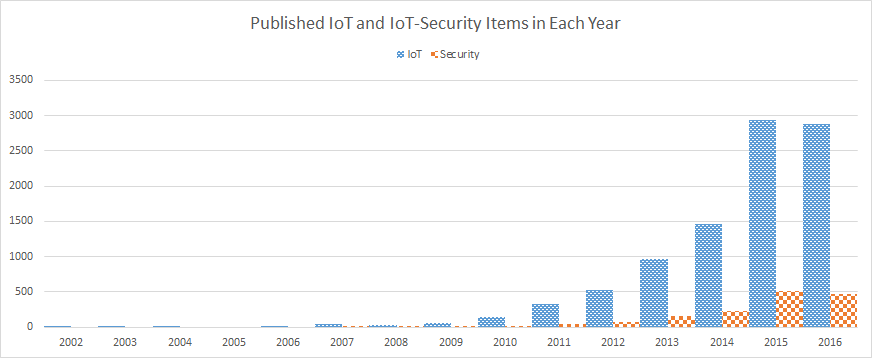}
\caption{\label{fig:IoTSecPubs}Number of publications for Internet-of-Things and Internet-of-Things Security related articles. As of December 2016 \cite{webscience2016db}.}
\end{figure*}

Security issues of IoT devices occur in different instances which include technological, ethical and privacy concerns. In October 2016, the massive Distributed Denial of Service (DDoS) attack on Dyn -  a company that controls much of the Internet’s domain name system (DNS) infrastructure - by a botnet army of IoT infected devices, has turned on the alarms on the consequences that faulty IoT protections and poor standards can motivate \cite{krebs2016iot} which accentuates the need for additional research on the IoT security domain. Nevertheless, the number of publications addressing security issues and concerns for the expanding Internet of Things has not foster the same attention to scientists in the community, even though the number of publications for IoT technologies and applications has grown exponentially during the last five years \cite{webscience2016db}. Figure \ref{fig:IoTSecPubs} shows a basic comparison of the number of publications for both subjects. 
In despite of the issues presented above, there is some important discussion taking place between experts about what the baseline for secure IoT systems. For instance, according to \cite{borgia2014internet}, IoT devices demand the following set of security requirements in order to be considered as secure : 
\begin{itemize}
\item Secure authentication
\item Secure bootstrapping and transmission of data
\item Security of IoT data
\item Secure access to data by authorized persons
\end{itemize} 
Weber \cite{weber2010internet}, has also determined similar security requirements for the Internet of Things, which include: (a) attack resiliency, (b) data authentication, (c) access control, and finally demand (d) client Privacy. Also, \cite{zhao2013survey} proposes security requirements to protect IoT data transmission, which include the following: (a) key management, (b) appropriate secret key algorithms, (c) secure routing protocols, (d) intrusion detection technology, (e) authentication and access control, and finally, (f) physical security design.\\ 
For \cite{zafari2015micro,zafari2016micro}, the two main security related issues have to do with data integrity and authentication. At the moment, the security and privacy requirements face serious challenges since current technologies do not offer feasible and comprehensive solutions applicable to the nature of the IoT. The unique scalability and distribution properties  of the Internet of Things call for flexible and innovative security frameworks that can close the existing gaps and reduce the risk associated with the use of embedded computing devices. The energy-efficient principle as well as the low computing properties of IoT devices are antagonistic to the essence of cryptography algorithms of current security protocols, determined as the "security processing gap" according to \cite{ukil2011embedded}. IoT devices are also exposed to physical tampering, war driving, malicious software and side-channel attacks \cite{ukil2011embedded}.\\
Security problems of the Internet of Things need to be understood in order to find an appropriate solution (Figure \ref{fig:IoTSecGraph}). The vulnerable landscape can be scrutinized from an architectural perspective; the perception, the network and the application layer present security problems of their own that need to be addressed as a whole. In section IV these layers are discussed in detail. Additionally, some other issues arise when the IoT platform is looked from a different technical perspective. For instance, \cite{fernandes2016security} states that 55\% of Samsung-owned SmartThings development platform applications are overprivileged and therefore present important security risks. \\

\section{Enabling Technologies and Protocols}
The Internet of Things (IoT) may be powered by different technologies with different properties for distinctive applications. However, those technologies also bring up some security issues that need to be addressed based on the capabilities and constrains that IoT devices offer at each IoT layer. This paper presents the following security concerns based on the IoT threat model presented by \cite{atamli2014threat}, and specifically related to the external adversary entity. The authors of \cite[p. 38]{atamli2014threat} refer to the external adversary as: ``An outside entity that is not part of the system and has no authorized access to it. An adversary would try to gain information about the user of the system for malicious purposes such as causing financial damage and undermining the user’s credibility. Also, causing malfunction to the system
by manipulating the sensing data''. 
\subsection{Perception Layer}

\subsubsection{Wireless Sensor Networks} %Added by Diego Mendez
The authors in \cite[p.65]{boyle2008securing} defined Wireless Sensor Networks (WSN) ``as a group of independent nodes communicating wirelessly offer limited frequency and bandwidth", which in order to perform successfully depend on a massive deployment and strict coordination. The limitations of WSN include ``power management, network discovery, control and routing, collaborative signal and information processing, tasking and queering, and security". According to \cite{gubbi2013internet} the WSN network module include the following components:
\begin{enumerate}[(a)]
\item Hardware
\item Communication stack
\item Middleware
\item Secure data Aggregation
\end{enumerate} 
Similar to active Radio Frequency Identification (RFID) technology, the data collected by the sensor nodes is shared between them or by a centralized system for analytic purposes \cite{gubbi2013internet}. \\
A WSN is composed of the following elements: 
\begin{enumerate}[(a)]
\item Sensor
\item Micro-controller
\item Memory
\item Radio transceiver
\item Battery
\end{enumerate}
A WSN consists of a centralize base station that controls a multi-hub relay system that connects the source nodes and the base \cite{borgohain2015survey}.
WSNs, as well as other network applications, require measures against common attacks and rises which include Denial of Service (DoS), traffic analysis, node replication (Sybil attack), general confidentiality concerns, black hole routing attacks and physical damage / unauthorized manipulation \cite{boyle2008securing}. \cite[p.66]{boyle2008securing} also mention the necessity of a "common communication protocol" in order to find a feasible solution for system protection at the application level which includes IEEE 802.15.4 Security, Zigbee and Tiny OS protocols. Moreover, some security requirements are set in order to consider WSN nodes as secure, which include:
\begin{itemize}
\item Data confidentiality
\item Integrity
\item Freshness
\item Availability
\item Organization autonomy
\item Authentication
\end{itemize}
WSNs vulnerabilities, according to \cite{borgohain2015survey}, can be categorized as the following: (a) attacks on secrecy and authentication, (b) silent attacks on service integrity and (c) attacks on network availability. Availability attacks (DoS) against WSN devises can occur on different layers of the network including DoS attacks on the physical layer (jamming, node tampering), DoS attacks on the link layer (collision, unfairness, battery exhaustion), DoS attacks on the network layer (spoofing, hello flood, homing, selective forwarding, sybil, wormhole, acknowledgement flooding), DoS attacks on the transport layer (flooding, de-synchronization) and DoS attacks on the application layer (traffic congestion generation). Attacks on WSN can further be classified on to one of the following categories: (a) external, (b) internal, (c) passive, (d) active, (e) mote-class, (f) laptop class, (g) interruption, (h) interception, (i) modification, (j) fabrication, (k) host-based and (l) network-based \cite{borgohain2015survey}.\\
Authors of \cite{sicari2015security} affirm that work has been done to secure WSN, however, some questioning has been arising. The questioning involves adaptability to the heterogeneous properties of IoT devices, network layer security management determination, feasibility of re-utilization of existing encryption protocols and end-to-end integrity verification. The authors also mention some additional efforts that include lightweight encryption methods, such as Elliptic Curve Cryptography (ECC),  to protect privacy and avoid counterfeiting attempts, which require additional standardization efforts to meet confidentiality expectations of the IoT infrastructure.  \\
WSN security concerns can be addressed in some aspects, by the use of authentication methods through Public Key Infrastructure (PKI), to prevent rogue node data injection, and authorization technologies to mitigate DoS risks \cite{medaglia2010overview}.\\
According to \cite{boyle2008securing}, node authentication can solve most of the problems that may be caused by unauthorized uses, some of the authentication methods disused take into account SPINS, composed of Secure Network Encryption Protocol (SNEP), micro-Tesla, TINYSEC, Localized Encryption and Authentication Protocol (LEAP/LEAP+) and Zigbee.

\subsubsection{Radio Frequency Identification} %Added by Diego Mendez
%Radio Frequency Identification (RFID) provides automatic identification, might be considered as an “electronic barcode”. Passive RFID tags do not need battery power, they use the power provided by the tag reader’s interrogation signal \cite{gubbi2013internet}. 
Radio Frequency identification (RFID) implementations provide unique identification based on passive tags to the items they are attached to. The data transmitted from the scan reading is commonly ``unprotected or read-only" \cite[p.391]{medaglia2010overview}, including Ultra High Frequency (UHF) and Global Gen-2 tags under default settings. RFID passive tagging, by default, permits reading by any compliant scanner with no authentication at all, increasing ears dropping risks and relegating passive RFID solutions to non critical settings \cite{medaglia2010overview}.
%RFID components interact between each other automatically. RFID transponders, also known as tags, are composed of embedded memory units hosting unique identifiers called Electronic Product Code (EPC). There are two classes of tags, passive and active; active tags need of an internal battery and the passive ones rely on the power inducted by electromagnetic signals emitted by a compliant RFID transceiver (reader) inside a specific working range \cite{borgohain2015survey}.
RFID vulnerabilities can be classified as the following: (a) attacks on authenticity, i.e. unauthorized tag disabling, (b) attacks on integrity, i.e. unauthorized tag cloning, (c) attacks on confidentiality, i.e. unauthorized tag tracking and (d) attacks on availability, i.e. replay attacks \cite{borgohain2015survey}. Also, Corporate espionage risks, location as well as personal privacy concerns may be affected by the use of unprotected tags \cite{weis2004security}. 
Usually, many RFID implementations are exposed to physical and traffic analysis attacks \cite{henrici2004tackling}, based on their autonomous properties, RFID devices respond automatically to readers' requests makes them inherently vulnerable \cite{kavun2010lightweight}. Back in 2004 and 2005, proof-of-concept attacks have been published against RFID financial transactions disclosures as well as cryptographic keys brute-forcing for widely deployed RFID tags \cite{phillips2005security}. Even Advanced Encrypted Scheme (AES) RFID solutions may present security vulnerabilities a those devices, based on the passive power capabilities, are susceptible to ``fault induction, timing attacks or power analysis attacks" \cite[p.203]{weis2004security}. \\
Zhao \& Ge \cite{zhao2013survey} provide security measures to secure RFID communications, which include: (a) access control, that protect RFID tags against reading at will, (b) data encryption, which include non-linear key algorithms, (c) IPSec protocol utilization, (d) cryptography technology scheme, to protect against side-channel attacks, i.e. Differential Power Analysis (DPA) attacks by eliminating "data dependencies of the energy consumption" and also by obscuring the encryption process values based on randomization\cite[p.666]{zhao2013survey}, however, DPA attacks are not protocol-dependent as it may be used against other communication procedures. \cite{kavun2010lightweight} has also presented a new lightweight implementation using the recent SHA-3 appointed function Keccak-f(200) and Keccak-f(400) to secure RFID applications through hashing functions. \cite{henrici2004tackling} proposes more simple solutions based on "tag killing", which is not satisfactory, and "Blinded Tree-Walking" to take advantage the backward channels to prevent eavesdropping attacks. Nevertheless, both approaches do not address location privacy issues. \cite{henrici2004tackling} also presents "Blocker Tags" solutions to provide user privacy by blocking readers' requests using the anti-collision algorithm embedded in RFID systems to differentiate between tags. Other solutions propose ciphertext re-encryption to hide communication appearance by using encryption-capable tags which usually do not offer a cost-effective solution \cite{henrici2004tackling}.
\cite{phillips2005security} promotes the development of security standards not only for emerging RFID technologies to address security issues but also to solve other interaction issues, which can be seen as a win-win situation for not-security-focused developers. \cite{weis2004security} also proposes a hashed based access control by using hardware-optimized functions as well as randomized access control to prevent user tracking by using hash functions and random number generators in order to control the predictability of responses from RFID tags.

\subsubsection{802.11} %added by Sahithya Kodam and Diego Mendez
The current IEEE 802.11 standard, 802.11ac or 802.11n i.e. Wi-Fi is wireless Local Area Networking technology that connects electronic devices within the range of 100m to the network. IEEE has developed 802.11ah Task Group that is focused on developing a wireless communication standard based on 802.11 which is suitable for the specific needs of IoT systems. 
The main intent of the task group is to eliminate the existing capability and capacity gaps. 
%The requirements set by the task group members Adame, Toni, et al are (a) Up to 8,191 devices associated with an access point (AP) through a hierarchical identifier structure, (b) Carrier frequencies of approximately 900 MHz (license-exempt) that are less congested and guarantee a long range, (c) Transmission range up to 1 km in outdoor areas, (d) Data rates of at least 100 kbps, (e) One-hop network topologies, (f) Short and infrequent data transmissions (data packet size approximately 100 bytes and packet inter-arrival time greater than 30 s.), (g) Very low energy consumption by adopting power saving strategies, (h) Cost-effective solution for network device manufacturers. 802.11ah is designed to have new PHY and MAC layers that include several modifications when compared to existing standards for supporting special constraints of IoT systems. This standard is still in its development phase. Further study of the protocol will be possible on its standardization. 
IEEE 802.11 networks are vulnerable, as any other wireless network, to easily performed passive attacks, including eavesdropping just armed with suitable receiving antennas \cite{djenouri2005survey}. However, active attacks can be performed as well by exploiting protocol and hardware vulnerabilities including jamming and scrambling attacks \cite{naeem2009common}.  
The current 802.11ah development and design requirements also try to implement security based on the properties of IoT devices, such as constrained memory and limited power supply, in order to include them in early stages to ensure proper functioning when fully deployed \cite{80211ahupdate}.

\subsubsection{Long Term Evolution (LTE)/LTE-Advanced} %added by Diego Mendez
IoT Systems depend on gateways to reach the Internet in an efficient way, for areas where the wired gateways are not an option usually Long Term Evolution (LTE) devices are chosen to fulfill that purpose based on bandwidth, coverage and spectrum efficiency \cite{costantino2012performance}. LTE Femtocells, which are used as low-range and low-power radio bases designed for small scale users or systems, provide the connection to the core cellular network. Such low-tier cells level the ground that eventually fosters the spread use of LTE and its significance \cite{bilogrevic2010security}. According to \cite{bilogrevic2010security}, ``Security and Privacy in such networks is achieved at several levels in their air architectures, such as the air interface, the operator's internal network and the inter-operator links" \cite[p. 1]{bilogrevic2010security}. As any other wireless technology, LTE networks are susceptible to passive and active attacks, although some active attacks can be controlled by the use of cryptographic tools. Passive attacks, such as traffic analysis and  accurate user tracking are nearly impossible to contain \cite{bilogrevic2010security}. Femtocells are also vulnerable to tampering as attackers may found them easier to access that any other LTE infrastructure, which can lead to undetected privacy exposure. Moreover, Femtocells are exposed to other kind of attacks, including impersonation, false reporting of location that may affect the normal operation of the device \cite{bilogrevic2010security}. 
Exposure of public IP addresses of gateways, such as Femtocells, could leave the LTE core network vulnerable to internet-originated attacks, such as DoS, DDoS and impersonation attacks as well \cite{bilogrevic2010security}.  
Some solutions proposed to address identity and location tracking is by the implementation of adaptive schemes that change the identifiable information based on the context of the communication or by demand of the user \cite{bilogrevic2010security}.

\subsubsection{WiMax} %added by Faheem Zafari and edited by Diego Mendez

\par While the technology has lost its popularity, it can still be used to connect different IoT devices particularly in metropolitan areas. The higher data rate accompanied by longer range can certainly facilitate different entities particularly in remote areas. IEEE 802.16 security specifications reside primarily within the MAC layer, such specifications reside on what is called a 'security' or 'privacy' sublayer \cite{papapanagiotou2009survey}, therefore, the physical layer remains mainly unprotected \cite{hasan2009security}.  Some of the security concerns associated with WiMAX are jamming at physical layer that can result in denial of service \cite{hasan2009security} or network mapping by eavesdropping \cite{bogdanoski2008ieee}. Nevertheless, the MAC layer also presents security issues such as Man-in-the-middle attacks, caused by rogue Base Station (BS) that pretends to be a legitimate BS, replay attacks and Denial or downgrade of service due to flawed authentication and resource limitation, which includes cryptographic computer efficiency constraints \cite{rengaraju2009analysis}.  \cite{huang2008responding} proposes the incorporation of additional schemes for authentication and key distribution that nevertheless still have efficiency issues to be improved before becoming introducing into real applications.%Interested readers are referred to \cite{hasan2009security,shon2007analysis,ahson2007wimax} for an in depth insight into security issues in WiMAX.  

\subsubsection{Near Field Communication} %added by Faheem Zafari
Near Field Communication (NFC) has a short range of 20 centimeters, however it can be used for wide range of services in IoT systems such as payments, authentication, data exchange, etc. From the security perspective, NFC is also prone to a number of threats including Denial of Service (DOS), and information leakage \cite{madlmayr2008nfc}.
The major security issue with NFC is that for some cases it is not encrypted, i.e. to maintain backward compatibility with RFID. Therefore, it results in security vulnerability as the wireless signal generated by the devices can be picked up by antennas \cite{curran2012near1}. NFC is also prone to eavesdropping in active mode. It is possible as well to implement an NFC skimmer device that could listen to the NFC communication between any two near-by devices. The data could be stored and collected later just like many ATM devices. It can also be manipulated by interfering with the data channel making the data corrupted and useless when it arrives at the destination. Similarly, the NFC tag can be modified by potential attackers who can replace the original tag with a fraudulent one with the intent to steal valuable user information. \cite{6490254} proposes a security model for NFC that provides conditional privacy protection. This method is based on the use of random public keys like pseudonyms. These keys are generated based on the long-term key issued by the Trusted Service Manager. This suggested method can protect user's identity and provide conditional privacy.

\subsubsection{Bluetooth} %Edited and Added by Diego Mendez
Bluetooth is certainly a viable technology for IoT systems. It has already been adopted for indoor proximity systems in form of iBeacons \cite{estimote}. Due to its range and data throughput, it can also be used in different sensor networks for various tasks such as earthquake monitoring. %Particularly in sensors which have no excess to power and rely on battery power, BLE is certainly a viable solution. %Sensors within a cluster, since they do not require high range, can exchange data or forward it to the cluster head using bluetooth. Similarly, intra-vehicular communication can also leverage bluetooth for providing in-car services to the driver. 
The Bluetooth protocol it is designed to provide security in 3 ways: (1) use of pseudo-random frequency hoping, (2) Restricted authentication and (3) Encryption \cite{bouhenguel2008bluetooth}. Even though the generic access protocol of Bluetooth make possible three security modes: (1) Non-Secure, (2) Service-levels security and (3) Link-level security, there are still security concerns that need to be addressed. \cite{sharma2008bluetooth} lists some vulnerabilities that persists even after the security features have been introduced, which includes optional or weak encryption, non-secure default settings, weak PIN use, insecure unit keys, flawed integrity protections and predictable number generation. Bluetooth is also prone to a number of threats including eavesdropping, Man-In-The-Middle attacks, data corruption, and denial of service. Attackers have also paid attention to vehicle IoT Bluetooth pairing applications and devices making them a valuable target that need to be secured \cite{oka2014survey, Bayram2014}. Nevertheless, \cite{bouhenguel2008bluetooth} proposes some simple solutions that address some of the security flaws, which include: User understanding of the technology, centralized Bluetooth pairing policy implementation, use of non-discoverable mode or on-demand access/pairing and mandatory encryption use. 
Bluetooth has also been used for microlocation purposes through beacons enabled by Bluetooth Low Energy (BLE) technology \cite{zafari2015enhancing,faheem2017icc} or proximity applications \cite{zafari2017arxiv}. BLE communication is conformed by the interchange of small data packets which are broadcast, one-way only, within an specific time \cite{dudhane2015location}. As expected, the data processed by beacons and BLE systems may contain private user data that need to be protected from intruders and from indiscriminate use \cite{dudhane2015location}. BLE uses AES-128 CCM for encryption and authentication purposes \cite{zafari2015micro}. 
%Unfortunately, the best solution that currently exists is turning off the Bluetooth when not in use. 

\subsubsection{ZigBee} %added by Sahithya Kodam and Diego Mendez
Based on the IEEE standard 802.15.4, the ZigBee protocol defines the network (NWK) layer as the one that runs on top of the physical (PHY) and MAC layers. It also comprehends descriptions, algorithms and protocols for the application support (APS), which outlines the device objects (ZDO), the device profile (ZDP) and the security services. The present security architecture consists of the ZigBeee coordinator which performs network joining and key distribution duties, here the Thrust Center concept was introduced, which plays three roles: 
\begin{itemize}
\item Trust manager
\item Network manager
\item Configuration manager
\end{itemize}
The trust manager authenticates the devices requesting to join the network, the network manager maintains and distributes network keys and the configuration manager provides end-to-end security between devices \cite{alliance2006zigbee}. ZigBee also offers two operation modes, residential and commercial, where the first one offers no security and the second one provides a centralized key management scheme and maintains freshness counters with other devices in the network. This gives centralized control and the ability to update keys. The three different types of keys are the Master key, Link Key and Network Key. The Master Key is part of the factory-setting. They are from the Trust Center and are the basis for long-term security between devices. The network Keys are shared by all the devices on the network and are the basis for security over the entire network. The Link Key is shared by two devices and is the basis for the security between the devices. The use of the different keys ensures the freshness and integrity of the data. ZigBee uses AES-128 with CCM (CCM = CBC-MAC = Counter with Cipher Block Chaining Message Authentication Code) mode for data encryption and authentication. However, some trade-offs must be made between security, power consumption and latency (Boyle \& Newe, 2008). According to \cite{vidgren2013security}, Zigbee-enabled systems are susceptible to threats such as  traffic sniffing (eavesdropping), packet decoding, and data manipulation/injection, which can be exploited by an attacker who uses  a special hardware and software developed especially for attacking purposes. Other vulnerabilities that a ZigBee enabled network is susceptible to are physical device damage and key sniffing attacks. \\
A similar proprietary approach is brought by Z-Wave, backed up initially by Intel and Cisco \cite{knight2006wireless}. Z-Wave has an architectural similarity with Zigbee with a few differences that makes Z-Wave a more approachable technology for home-automation systems. According to \cite{fouladi2013security}, Z-Wave uses AES encryption for secure communication, however, it presents two main vulnerabilities. The first one deal with the hard-coded encryption key embedded in the Z-Wave chip, an attacker may intercept the key exchange messages and use the hard-coded key to unveil the content. The second one deals with the lack of validation of the key exchange protocol handler which can allow an attacker to inject packets to the key exchange process and take control of the device. A security feature, noted by \cite{knight2006wireless}, includes the procedure to add new devices to the network that forces the new device to be within one meter for set-up, which limits the attack ratio. Additionally, as a response to recent hijacking of IoT devices the Z-Wave Alliance has come up with the security framework S2 \cite{zwaves22017}. The alliance, fortunately, has made the implementation of the new framework mandatory and, therefore, improved the embedded security by strengthening the pairing process for new devices \cite{zwaveall2017}. 

\subsubsection{Ultra-Wideband} %added by Faheem Zafari
%Systems that occupy spectrum in excess of 500MHz are based on  Ultra-Wideband (UWB) technology \cite{yang2004ultra}. Due to such high frequency, the waveforms for UWB are very small and do not usually include the sine wave carriers. UWB can operate at baseband and does not require any Inverse Fourier (IF) processing. Due to higher frequency and shorter duration, UWB has the following advantages \cite{yang2004ultra}:
%\begin{enumerate}
%	\item Penetrate efficiently through obstacles
%	\item Provide higher throughput
%	\item Higher network capacity
%	\item Higher precision ranging i.e. within centimeters
%	\item Small size and improved processing power
%\end{enumerate}
%Prior to 2001, UWB's use was restricted only to radar systems and was primarily used by the military. However, in 2002, the FCC allowed utilization of UWB at noise floor and huge bandwidth ranging as high as 7.5GHz. This allows UWB to be used for ultra-low power, position critical applications such as localization, imaging, radar etc. 
The low power and higher precision make it suitable for the IoT and smart architecture applications as well including the provision of micro-location services for tenants of IoT-equipped smart buildings. %UWB can provide data rates as high as 110 MB/s at very low power and a short range of about 15 meters. As of now, the technology faces a number of challenges including the network capacity, Bit Error Rate (BER), flexibility and throughput. UWB is highly sensitive and requires high synchronization of the received ultra-short pulses. Furthermore, there is also a need for optimally exploiting the fading propagation effects. \par
From security perspective, literature shows the UWB is comparatively secure and is suitable for IoT applications due to its processing power, range, data rate and security. Indeed, it is already considered as a viable technology for Wireless Body Area Networks (WBANs) and can be used for a number of different services including smart health care system. UWB signals secure and highly reliable particularly in health applications \cite{ullah2009applications}.

%The Ultra-Wideband (UWB) technology has a fractional bandwidth which is at least  20\% where fractional bandwidth is the ratio of transmission bandwidth to the band center frequency  \cite{gezici2005localization}. UWB has an absolute bandwidth greater than 500 MHz. There are a number of advantages associated with using high bandwidth that can facilitate both communications and radars based applications. Using high bandwidth provides higher reliability because the probability of signals going around any obstacle increases due to the availability of wide range of signals having different frequencies. Also the power spectral density decreases since the signal power is spread over a large number of frequencies. Also there is a decrease in interference as well as interception probability. 

\subsubsection{IPv6 Low Power Personal Area Networks (6LoWPAN)} %added by Sahithya Kodam
Since the conceptualization of IoT technologies, research has inclined to select IPv6 as the choice for wireless communication. 6LoWPAN communication standard applies IPv6 to the PHY and MAC layer of the existing 802.15 standard. According to Sheng, Zhengguo, et al. \cite{sheng2013survey} the key features of IPv6 which makes it suitable for the IoT are universality, extensibility, and stability. It has special characteristics such as small packet size, low bandwidth, and large number of devices. According to Park, S., et al. \cite{park2011ipv6} the security challenges for a 6LoWPAN network are (a) minimizing resource consumption and maximizing security
Performance, (b) 6LoWPAN deployment enables link attacks ranging from passive eavesdropping to active interfering, (c) in-network processing involves intermediate nodes in end-to-end information transfer, (d) 6LoWPAN communication characteristics render traditional wired based security schemes unsuitable. 6LoWPAN is susceptible to various attacks, the list of threats based on ISO OSI layers: (a) 6LoWPAN devices are vulnerable to physical attacks like node tampering, destruction and masking. Several types of DoS attacks can be triggered at different layers. At physical layer, jamming and and node tampering. (b) Attacks at MAC layer include collision, battery exhaustion and unfairness. (c) At network layer, 6LoWPAN is vulnerable to spoofing attacks as well as altered, or replayed routing information attacks, selective forwarding, sinkhole attack, Sybil attack, wormhole attack, neighbor discovery attacks. (d) An attack against the transport layer is performed by half open, half closed TCP segment. The attacker continuously forges messages carrying sequence numbers or control flags. This will cause the endpoints to request retransmission of missed frames leading to DoS attack due to large amount of traffic. 
A secure 6LoWPAN protocol should provide:
\begin{itemize}
\item Data confidentiality
\item Data authentication
\item Data integrity
\item Data freshness
\item Availability
\item Robustness
\item Resiliency
\item Resistance
\item Energy efficiency
\item Assurance
\end{itemize}
To ensure maximum security 6LoWPAN should employ secure bootstrapping mechanisms, Secure Neighbor Discovery protocol (SeND) extended to support Elliptic Curve Cryptography (ECC) encryption algorithm which uses smaller-packet sizes compared to RSA and secure key management algorithms engineered to suit the specific characteristics of 6LoWPAN.

\subsection{Middleware}
The challenges presented by the IoT can present issues between each one of the architectural components of embedded systems, therefore, middleware has been developed in order to interconnect and integrate all the elements that make the IoT possible. Middleware in the IoT is used as well to interact with ``cloud technologies, centralized overlays, or peer to peer systems".  Evidently, the attack surface increasing the demand for more comprehensive IoT security, moreover, the lack of standardized approaches do not permit a comprehensive response to all IoT security and privacy requirements. Services such as context-awareness may risk personal privacy as critical user information may be disclosed by malicious parties \cite{razzaque2016middleware}.  The authors of \cite[p.76]{razzaque2016middleware} proposes seven categories for discussion based on design principles:

\subsubsection{Event-based}
According to \cite{razzaque2016middleware} all the participants in the middleware connect through events, the events consist of a set of parametric values that describe specific changes of state. Some event-based middleware applications present security features and some others do not consider security requirements at all. For instance, HERMES \cite{pietzuch2004hermes} utilizes a security module that controls the perimeter based on access control, it also provides confidentiality between brokers through X.509 certificates and OASIS role memberships. Other applications such as EMMA, GREEN, RUNES, Steam, MiSense, PSWare and TinyDDS do not show specific security features \cite{razzaque2016middleware}.   

\subsubsection{Service-oriented}
\cite{razzaque2016middleware} describes service-oriented middleware to the design approach that constructs applications as services, similar to service-oriented computer (SOC) that is based on service-oriented architecture (SOA) for common Information Technology (IT) systems. Service-oriented applications include security attributes as well as vulnerabilities. HYDRA \cite{eisenhauer2010hydra} employs a security manager as part of its management components design, each one of the components take care of application and device elements, which have an additional security layer. It applies a distributed security as well as trust elements for securing inter-device connections, it also uses virtualization to provide security and privacy, which according to \cite{razzaque2016middleware} may introduce vulnerabilities for side-channel attacks. SOCRADES \cite{guinard2010interacting}, introduces role-based access control for device communication with the application, however, its security features is limited to authentication only. UbiSOAP \cite{caporuscio2012ubisoap} uses some functions from its resource layer to authenticate components for security and privacy. Servilla \cite{fok2012servilla} presents privacy concerns due to the access level it provides to individual sensors\cite{razzaque2016middleware}. KASOM \cite{corredor2012knowledge}, uses a security manager which is part of one of the major subsystems, however, it offers security by authentication only \cite{razzaque2016middleware}. Xively \cite{xively2017site} provides support for end-to-end security for the entire system which provides integrity, however, it does not provide security for its storage components \cite{satyadevan2015security} same for CarrIoTs \cite{Carriots2017site}.  Echelon \cite{Echelon2017site}, according to \cite{razzaque2016middleware}, does not include any security mechanism. Other applications, such as SenseWrap, MUSIC, TinySOA, SensorsMW, SENSEI, KASOM, CHOReOS, MOSDEN and WhereX do not show specific security components for analysis \cite{razzaque2016middleware}. 

\subsubsection{Virtual Machine (VM) based}
\cite{razzaque2016middleware} classifies middleware as VM-based the applications that use virtual infrastructure for the purpose of safe execution. The applications are built from specialized modules, which can be spread through the network, where each node runs a VM that interact with the modules. The middleware application Mat\'e \cite{costa2007virtual} has a system component dedicated to security, its purpose is to block the propagation of harmful programs through the network. Other applications, such as VM\*, Melete, MagnetOS, Squawk, Sensorware, DVM, DAViM, SwissQM and TinyReef do not show any specific security features for analysis according to \cite{razzaque2016middleware}.

\subsubsection{Agent-based}
Agent-based middleware is composed of modular programs that ``facilitate injections and distribution through the network using mobile agents". Agent-based middleware deserve as well a security-focused analysis for features and vulnerabilities. The application Ubiware \cite{nagy2009challenges} adds policies to support the security requirements of the middleware. Other solutions, such as Impala, Smart Messages, ActorNet, Agilla, UbiROAD, AFME, MAPS, MASPOT, and TinyMAPS, do not present security or privacy demands \cite{razzaque2016middleware}.

\subsubsection{Tuple-spaces}
Tuple-space middleware each component contains a data repository or tuple space that can be accessed simultaneously \cite{razzaque2016middleware}. Applications such as LIME \cite{murphy2001lime}, TinyLIME \cite{curino2005mobile} and TS-Mid \cite{lima2008ts} do not support any security or privacy mechanism \cite{razzaque2016middleware}. 

\subsubsection{Database-oriented}
The sensor network acts as a ``virtual relational database system" that can be queried by SQL-alike language. Under the security domain, current database-oriented middleware solutions, such as GSN \cite{aberer2006middleware} and HyCache \cite{zhao2013hycache} do not deal with security or privacy requirements. TinyDB \cite{tinydb2017site} does not show any security feature for analysis according to \cite{razzaque2016middleware}.

\subsubsection{Application-specific}
Application-specific middleware specializes on managing resources for specific requirements demanded by the application or by the domain it works on \cite{razzaque2016middleware}. Given the nature of application-specific middleware the solutions developed under these architecture cannot satisfy general IoT requirements taken into account in other applications in terms of heterogeneity, therefore, security demands cannot be satisfied integrally \cite{razzaque2016middleware}.    

\subsection{Application Layer}

\subsubsection{Message Queue Telemetry Transport (MQTT)} %added by Sahithya Kodam edited by Diego Mendez
The characteristics of the various devices used in Internet of Things is such that they lack the capability to handle high-level protocols like HTTP. Researchers are more inclined on developing light-weight protocols that suit the specific characteristics of IoT devices. The Message Queue Telemetry Transport (MQTT) proposed by Andy Stanford-Clark, and Arlen Nipper \cite{stanford2014mqtt} in 1999 is a light-weight protocol designed for constrained devices and low-bandwidth, high-latency or unreliable networks. %It is fast, power-efficient and employs various levels of Quality of Services. MQTT is based on a publish/subscribe (pub/sub) pattern with a central broker. Clients in MQTT exchange messages through the broker by publishing or subscribing to a topic. MQTT libraries have been developed for major IoT platforms, like Arduino, for several programming languages like C, Java, Python, Javascript and for mobile platforms, Android and iOS. 
The present implementation of MQTT provides support for only identity, authentication and authorization policies. Identity specifies the client that is being authorized. Authentication provides identity of the client and authorization is the management of rights given to the client. The basic approaches used to support these policies are by using a username/password pair, which is set by the client, for identification or by authentication performed by the MQTT server via client certificate validation through the SSL protocol. The MQTT server identifies itself with its IP address and digital certificate. The MQTT communication uses TCP as transport layer protocol. By itself the MQTT protocol does not provide encrypted communication. Authorization is also not part of MQTT protocol. Authorization is provided by MQTT servers. MQTT authorization rules control which client can connect to server and what topics a client can publish or subscribe to. 
According to Neisse \cite{neisse2014enforcement} the security controls provided by MQTT are not sufficient for the IoT network. IoT networks requires ``data anonymization, obfuscation or dynamic context-based policies that should be dynamically evaluated for each message forwarded by the broker" \cite[p. 1]{neisse2014enforcement}. Neisse \cite{neisse2014enforcement} implements a solution for the enforcement of security at MQTT layer which is a Model-based Security Toolkit called SecKit. It addresses the privacy and data protection requirement. For secure communication, security mechanisms have to be adopted over existing MQTT protocol. \cite{singh2015secure} proposes a new security solution for MQTT (Secure MQTT or SMQTT) that replaces the use of SSL/TLS certificates, which are not necessarily viable in all IoT implementations, the solution runs over Lightweight Attribute Based Encryption (ABE) over elliptic curves.

\subsubsection{Extensible Messaging and Presence Protocol (XMPP)} %added by Sahithya Kodam
The Extensible Messaging and Presence Protocol is an application profile of the Extensible Markup Language (XML) that enables the near-real-time exchange of structured and extensible data between any two or more network entities. The core features of XMPP provide the building blocks for different types of near-real-time applications, which can be layered on top of the core by sending application-specific data qualified by particular XML namespaces\cite{saint2011extensible}. XMPP architecture is defined by a distributed network of clients and servers. The recommended ordering of layers in XMPP described in \cite{saint2011extensible}, in order to ensure security is to have TCP, followed by TLS, SASL and then XMPP. Using XMPP over TLS provides confidentiality and integrity to data which is in motion over the network. Unless the network is protected with TLS, it is open to attacks. But the XMPP protocol does not provide end-to-end security. The data is processed in cleartext on the sender's and the receiver's servers. It is also unprotected when it is sent from sender's to receiver's server, or sent from receiver's server to receiver's client. Systems using XMPP as the enabling technology must ensure that they use secure protocols along with XMPP. For authentication purposes, the servers and the clients should support Salted Challenge Response Authentication Mechanism (SCRAM). Using both TLS and SCRAM provides both confidentiality and authentication. Due to its capability of real-time message exchange, XMPP is a viable enabling technology for the IoT but XMPP has to be used in conjunction with the various security protocols to ensure confidentiality, integrity and authentication of the IoT system.         

\subsubsection{Blockchain}    % added by YANG 5/10/17
Blockchain \cite{christidis2016blockchains} was originally proposed in Bitcoin to solve the double spending problem in a cryptocurrency system. However, a blockchain can stand by itself and be applied in a distributed and trustless environment without the need of third party authentication or management. In a nutshell, a blockchain is a back-ordered hash list that is publicly shared in a peer-to-peer network. Usually, each member in the blockchain system is addressable by the hash value of its public key. When a new transaction occurs, the owner of the transaction can prove the authenticity of the record (i.e. block) by encrypting the hash value of the record using its private key. The newly formed block is then appended to the existing blockchain and point to the previous block. Supported by the cryptographic properties of hash and asymmetric encryptions, a blockchain can therefore ensure each block is immutable and transaction is verifiable. Blockchains has recently received a lot of attentions in the field of IoT. Researchers and practitioners believe blockchain is one of the key technologies that can securely enable smart contracts among the ‘things’. That is, smart devices can interact and transact with each other autonomously without human interventions. Though it is possible to implement blockchains in a public network, the computing overhead of providing proof of work (mining) may overwhelm the limited computing resources in an IoT network. If on the other hand, participating members in a blockchain network are not completely trustless, simple techniques, such as whitelisting, can be leveraged to reduce the burden of mining and make blockchains much more desirable in real world practice. It should be noted that, blockchains offer only pseudo anonymity: it is possible for adversaries to make inferences about who owns what public keys. If privacy is a major concern in an IoT system, additional mechanism must be designed and implemented to prevent the owners of the smart devices being identified.

\begin{table*}[!ht]
\centering
\caption{\label{tab:comparison}Comparison between most cited publications for IoT Security \cite{webscience2016db}}
\begin{tabular}{llllllllllll}
                                         &                           & \begin{turn}{80}\textbf{Gubbi et al. (2013)\cite{gubbi2013internet}}\end{turn}   & \begin{turn}{80}\textbf{Miorandi et al. (2012)\cite{miorandi2012internet}}\end{turn} & \begin{turn}{80}\textbf{Bandyopadhyay et al. (2010)\cite{bandyopadhyay2011internet}}\end{turn} & \begin{turn}{80}\textbf{Roman et al. (2011)\cite{roman2011securing}}\end{turn}   & \begin{turn}{80}\textbf{Zhou et al. (2011)\cite{zhou2011multimedia}}\end{turn}    & \begin{turn}{80}\textbf{Hong et al. (2010)\cite{hong2010snail}}\end{turn}    & \begin{turn}{80}\textbf{Zuehlke (2010)\cite{zuehlke2010smartfactory}}\end{turn}        & \begin{turn}{80}\textbf{Sarma et al. (2009)\cite{sarma2009identities}}\end{turn}   & \begin{turn}{80}\textbf{Roman et al. (2011)\cite{roman2011key}}\end{turn}   & \begin{turn}{80}\textbf{Yan et al. (2014)\cite{yan2014survey}}\end{turn}     \\ \midrule
\multirow{3}{*}{} \multirow{3}{*} {\textbf{Enabling Technologies}}  & \textit{WSN}              & \multicolumn{1}{c}{\checkmark} &                                 &                                      & \multicolumn{1}{c}{\checkmark} &                                & \multicolumn{1}{c}{\checkmark} &                                &                                & \multicolumn{1}{c}{\checkmark} & \multicolumn{1}{c}{\checkmark} \\ \cmidrule(l){2-12} 
                                         & \textit{RFID}             & \multicolumn{1}{c}{\checkmark} &                                 &                                      & \multicolumn{1}{c}{\checkmark} & \multicolumn{1}{c}{\checkmark} &                                & \multicolumn{1}{c}{\checkmark} &                                & \multicolumn{1}{c}{\checkmark} & \multicolumn{1}{c}{\checkmark} \\ \cmidrule(l){2-12}  
                                         & \textit{Ultra-Wide Band}     &  &   &  &  &                                &                                &  &                                &  &  \\ \midrule 
                                          
      & \textit{802.11}           &                                &                                 &                                      &                                &                                &                                &                                &                                &                                &                                \\ \cmidrule(l){2-12} 
\multirow{5}{*}{\textbf{Low-Level Protocols}}                                         & \textit{LTE/LTE advanced} &                                &                                 &                                      &                                &                                &                                &                                &                                &                                &                                \\ \cmidrule(l){2-12} 
                                         & \textit{NFC}              &                                &                                 & \multicolumn{1}{c}{\checkmark}       &                                &                                &                                &                                &                                &                                &                                \\ \cmidrule(l){2-12} 
                                         & \textit{Bluetooth}        &                                &                                 &                                      &                                &                                &                                &                                &                                &                                &                                \\ \cmidrule(l){2-12} 
                                         & \textit{6LoWPAN}          &                                &                                 &                                      &                                &                                & \multicolumn{1}{c}{\checkmark} &  &  & \multicolumn{1}{c}{\checkmark} & \multicolumn{1}{c}{\checkmark}
                                         \\ \cmidrule(l){2-12} 
                                         & \textit{ZigBee}          &                                &                                 &                                      &                                &                                &\multicolumn{1}{c}{\checkmark} & \multicolumn{1}{c}{}           &                                &  &  \\ \midrule\multirow{3}{*}{\textbf{High-Level Protocols}}
 & \textit{MQTT}  &  &  &  &  &  &  &                                &                                &  &  \\ \cmidrule(l){2-12} 
                                         & \textit{XMPP}        &  &   &                                      &  &  &  &                                &                                &  &  \\ \cmidrule(l){2-12} 
                                         & \textit{Blockchain}        &  &   &                                      &  &  &  &                                &                                &  &  \\ \midrule
\multirow{3}{*}{\textbf{Security Triad}} & \textit{Confidentiality}  & \multicolumn{1}{c}{\checkmark} & \multicolumn{1}{c}{\checkmark}  & \multicolumn{1}{c}{\checkmark}       & \multicolumn{1}{c}{\checkmark} & \multicolumn{1}{c}{\checkmark} & \multicolumn{1}{c}{\checkmark} &                                &                                & \multicolumn{1}{c}{\checkmark} & \multicolumn{1}{c}{\checkmark} \\ \cmidrule(l){2-12} 
                                         & \textit{Integrity}        & \multicolumn{1}{c}{\checkmark} & \multicolumn{1}{c}{\checkmark}  &                                      & \multicolumn{1}{c}{\checkmark} & \multicolumn{1}{c}{\checkmark} & \multicolumn{1}{c}{\checkmark} &                                &                                & \multicolumn{1}{c}{\checkmark} & \multicolumn{1}{c}{\checkmark} \\ \cmidrule(l){2-12} 
                                         & \textit{Availability}     & \multicolumn{1}{c}{\checkmark} & \multicolumn{1}{c}{\checkmark}  & \multicolumn{1}{c}{\checkmark}       & \multicolumn{1}{c}{\checkmark} &                                &                                & \multicolumn{1}{c}{\checkmark} &                                & \multicolumn{1}{c}{\checkmark} & \multicolumn{1}{c}{\checkmark} \\ \midrule                                         
\multirow{3}{*}{\textbf{AAA}}            & \textit{Authorization}    & \multicolumn{1}{c}{}           & \multicolumn{1}{c}{\checkmark}  &                                      & \multicolumn{1}{c}{\checkmark} & \multicolumn{1}{c}{\checkmark} &                                &                                & \multicolumn{1}{c}{\checkmark} & \multicolumn{1}{c}{\checkmark} & \multicolumn{1}{c}{\checkmark} \\ \cmidrule(l){2-12} 
                                         & \textit{Authentication}   & \multicolumn{1}{c}{\checkmark} & \multicolumn{1}{c}{\checkmark}  & \multicolumn{1}{c}{\checkmark}       & \multicolumn{1}{c}{\checkmark} & \multicolumn{1}{c}{\checkmark} &                                & \multicolumn{1}{c}{\checkmark} & \multicolumn{1}{c}{\checkmark} & \multicolumn{1}{c}{\checkmark} & \multicolumn{1}{c}{\checkmark} \\ \cmidrule(l){2-12} 
                                         & \textit{Accounting}       &                                &                                 &                                      &                                &                                &                                &                                & \multicolumn{1}{c}{\checkmark} &                                &                                \\ \midrule
                                         & \textit{Privacy}          & \multicolumn{1}{c}{\checkmark} & \multicolumn{1}{c}{\checkmark}  &   \multicolumn{1}{c}{\checkmark}                                   &    \multicolumn{1}{c}{\checkmark}                            &                                &                                &   \multicolumn{1}{c}{\checkmark}                             &        \multicolumn{1}{c}{\checkmark}                        &     \multicolumn{1}{c}{\checkmark}                           &       \multicolumn{1}{c}{\checkmark}                         \\ \bottomrule
\end{tabular}
\end{table*}

\section{Confidentiality, Integrity, Availability and Privacy Concerns for IoT Systems} %Added by Diego Mendez
An upcoming global network of ``things" brings challenges regarding security and privacy. Confidentiality, Integrity and Availability becomes paramount when exchanging data between IoT devices, the intelligence and autonomy of these devices demand further responsibility when protecting against device corruption and its influence in the network \cite{mayer2009security}. Different cryptographic and process-based solutions are available to assure and to provision confidentiality, integrity and availability, nevertheless, IoT systems demand not only these services, but also need to focus on how these solutions are executed and optimized \cite{heer2011security1},\cite{Singla_milcom15},\cite{Singla_cerias2015,yavuz_17}.  \\
The IoT relies heavily on wireless networks which are known to be vulnerable to all type of intrusions including unauthorized router access, faulty configurations, jamming, man-in-the-middle attacks, interference, spoofing, Denial of Service attacks, brute-force attacks, traffic injections, etc \cite{acharya2008data}. According to \cite[p. 14]{gubbi2013internet}, security is a main concern for large networks, therefore, IoT physical components are vulnerable to availability, confidentiality and integrity attacks. The ``first line of defense" is the application of cryptographic features. Encryption schemes protect confidentiality as message authentication codes assures integrity as well as authenticity. Former WSN implementations, according to \cite{christin2009wireless}, used to deal with attack models that required physical access to the nodes. Eventually, after opening WSNs to the Internet the threat model changed as attackers can reach WSNs ubiquitously where sensor nodes are the most vulnerable due to scarce computational resources.
According to \cite{uckelmann2011architectural}, in order to enable massive adoption of Internet of Things devices; security, including confidentiality, integrity, availability, and privacy issues must be addressed in order to make them trustworthy to the public.\cite{uckelmann2011architectural} suggests as well that there should exist different security levels since the requirements are not the same between devices. 
User privacy and integrity can also be endangered from the lack of data confidentiality and integrity. Unauthorized access of sensor data could interfere with the proper functioning of the system, as well as unauthorized access and control \cite{medaglia2010overview}. \\ 
IEEE standard 802.15.4, which provides guidelines to protect physical and medium access control layers, may be used as a instrument to add security features that sum up confidentiality, integrity and availability properties to the system \cite{medaglia2010overview}. The Internet Draft ID-Tsao \cite{tsao2011security} signals a high-level presentation of the existing threats and security countermeasures in terms of the security triad. \cite{garcia2012security} indicates a potential limitation of the framework based on the non-differentiation arguments and layer 3-only analysis. 
\subsection{Confidentiality}
\cite[p.1505]{miorandi2012internet} defines data confidentiality as a ``fundamental issue" for IoT solutions, ``particularly relevant in the business context". \cite{miorandi2012internet} also indicates that current data confidentiality solutions may not be applicable as is due to two main limitations: Amount of data generated and the effectiveness in the control of access to data of dynamic data streams. The authors in \cite{miorandi2012internet} also mention proper identity management as a key factor to assure data confidentiality. Some IoT devices need to handle data need to be classified as confidential. Confidentiality, of the communications channel, can be obtained through encryption schemes. Current symmetric and asymmetric algorithms should be analyzed before implemented based on the application, capability and the criticality of the IoT system as stated by \cite{alam2011interoperability}. \\
Wireless communications of things may be vulnerable to eavesdropping attacks that may compromise the confidentiality of the communication which could impact the node or the network as a whole \cite{garcia2012security}. \cite{suo2012security} emphasizes on the importance of confidentiality research and the inherent challenges attached to it. \cite{suo2012security} also indicates the importance of authenticity and the integrity of data groundwork as well. The confidentiality needed for a sensor data, according to \cite{suo2012security}, is not as important as the integrity and authenticity since the attacker may obtain the same values just by placing a rogue sensor next to the legitimate one. \cite{mayer2009security} states that the major confidentiality sensitivity, in the IoT context, resides in the communication, storage, localization/tracking, and identification. On the other hand, sensors, actuators, devices and processing topics are not as sensitive as the one listed in the first place. \cite{mayer2009security} believes that there is adequate research work available for communication, storage, localization/tracking, and identification subjects, although, \cite{mayer2009security} recognizes the complexity of employing current mechanisms for securing the IoT.  \\
According to \cite[p. 4]{mineraud2015gap}, IoT solutions must use security mechanisms that permit, based on the end-user decision, access to a ``predefined set of resources", also called data ownership.  It is needed then a differentiation for the security requirements of things based on criticality, \cite{heer2011security} indicates the importance on the difference for each of the IoT layers, the link layer, the network layer and the application layer. Current IoT technologies manage data security processes, including key management, which places a burden on IoT resources that may diminish IoT capabilities and increase risk \cite{singh2015twenty}. \cite{babar2011proposed} proposes the use of lightweight cryptographic algorithms so that the resource-limited IoT devices, especially for processing and storage capabilities, can provide data protection and, therefore, confidentiality. Datagram Transport Layer Security (DTLS) may be used as a solution to confidentiality problems by providing end-to-end security for the application layer. DTLS properties may also reduce the impact and the cost of resources of constrained devices compared to other solutions \cite{garcia2012security}. In order to protect data and communication confidentiality, some cloud-based solutions  establish secure channels based on cryptographic features relying on Public Key Infrastructure (PKI). Information flow control is noted as another solution to protect IoT data sharing utilizing a cloud-based platform that protects critical information as described by \cite{singh2015twenty}. 

According to \cite{bandyopadhyay2011internet}, privacy of humans and business confidentiality are two major issues to be addressed for the Internet of Things. Standard encryption schemes may solve the problem, however, energy and processing resources need to be efficiently applied, including key distribution mechanisms, to be considered as a valid IoT solution. Confidentiality and Privacy are usually tied together, according to \cite[p. 622]{weber2015internet}, ``Privacy as confidentiality represents solutions for anonymizing the collected data (including communications) and minimizing the collection of data". \cite{weber2015internet} also suggests that anonymous communication, such as hiding location, identity, time, frequency and volume details, as well as communication context is necessary to entirely protect traffic data from unauthorized access. \cite{weber2015internet} also indicates that in order to increase confidentiality levels it is important to apply privacy-base designs and privacy-enhancing technologies to make it possible. \\ 
Confidentiality also deals with government regulations and laws that demand data protection and confidentiality \cite{singh2015twenty}. \cite{coetzee2011internet} states that trust is fundamental for users of the IoT as the information shared by the things and the users will not be compromised. To do so, the principles of data confidentiality and security itself must be preserved. According to \cite{miorandi2012internet}, data confidentiality, privacy and trust are key factors that can leverage the widespread adoption of IoT technologies and applications. 
\subsection{Integrity}
IoT integrity deals with physical failures and damages at first sight, Integrity protection includes preservation against sabotage and the use of counterfeit units or components \cite{sadeghi2015security}. Another critical factor that influences data integrity is the robustness and fault tolerance capabilities of the IoT System \cite{miorandi2012internet}. Sensor networks, such as RFID solutions, face also other issues that limit their capability to overcome integrity problems as many of their components spend most of the time without being attended. Attackers can either modify the data while it is stored in the node or when it travels through the network. Read and write protections as well as authentication methods are common solutions to these issues. Data integrity is also ensured by password-based solutions, which brings into account the shortcomings of password protection, such as vulnerabilities related to password length and randomness. Also, the resources found in common IoT systems do not support typical cryptographic solutions because of the limited resources available \cite{atzori2010internet}. \\
Integrity for the Internet of Things not only is required to be guarded from external sources but also for internal processes, such as service integrity. Operating systems rigid process separation, known as Multi Level Security (MLS), help devices to avoid unauthorized modification from code running with high privileges. Nevertheless, MLS approaches have not been deployed widely as in some cases can be considered as expensive as well as not compatible with other IoT software. Other approaches to guarantee integrity use hash values which are stored externally to avoid compromises \cite{fongen2012identity}. Hardware solutions have also been proposed for integrity purposes, a challenge-based solution is mentioned in \cite{fongen2012identity} by the use of symmetric or asymmetric keys known as Trusted Platform Module (TPM). \\
Process integrity is also required by IoT devices, process integrity relies on the device, communication, and algorithm implementation integrity. The processing data correctiveness is highly desired to perform data processing for higher services and data correlation \cite{mayer2009security}.
Software integrity relies mostly on hardware isolation of critical code and data from other, less relevant, internal components and it can be hardware-enforced. SMART, SPM, SANCUS and Trustlite are some solution examples applied to devices with limited capabilities in terms of processing, power and battery life \cite{sadeghi2015security}. Nevertheless, hardware-based attacks, such as fault attacks, can compromise the integrity if the system does not have protections in place, i.e. perturbation sensors \cite{van2014encyclopedia}.   
Integrity verification for software configuration, called attestation, prevents malicious modifications and it is usually performed through secure hardware. However, IoT devices are forced to depend on software attestation which is based on ``strong assumptions" that are not easy to accomplish in practice. Instead, low-end embedded devices may use ``swarn attestation" that allows software integrity verification collectively from multiple devices or ``provers" \cite{sadeghi2015security}. \\ 
Authentication schemes used in the IoT not only tries to assure the identity of an object but also attempts to assure its integrity. Authentication, through Identity Management (IdM) provides resource control and helps to deliver auditing, accounting and access control as well. Nevertheless, the implementation of IdM presents some challenges when deployed in a IoT infrastructure after facing scalability, capability and management issues \cite{fongen2012identity}.
Standardized procedures are also important for ensuring integrity and quality as well. A common scheme permits the process development to satisfy data trustworthiness and traceability needs. Extensive collaboration between different IoT institutions and alliances is fundamental \cite{miorandi2012internet}.
\subsection{Availability}
Usual information networks, according to \cite[p. 648]{suo2012security}, need to guarantee ``identification, confidentiality, integrality and undeniability", nevertheless, IoT networks that can potentially be utilized in ``crucial areas of national economy", which need as well to pay special attention to availability and dependability. According to \cite[p. 601]{kasinathan2013denial}, for information networks, device availability is the ``most important factor".
IoT availability requirements, as specified by \cite{roman2013features}, are highly tied to reliability requirements. IoT systems need to display sufficient resiliency to sustain availability under desired levels as well as they need to guarantee a certain level of performance requested by their applications. Availability may also refer to ubiquitous requirements, \cite{al2015internet} proposes that in order for IoT devices to reach their potential they will need to address requirements, such as availability, which, by the way, it is listed as a one of the key challenges to be addressed for the IoT. \cite[p. 2362]{al2015internet} also indicates that the availability of the IoT networks should be performed in hardware and software so they can cope with user requirements. ``Availability of software refers to the ability of the IoT applications to provide services for everyone at different places simultaneously. Hardware availability refers to the existence of devices all the time that are comparable with the IoT functionality and protocols". \\
According to \cite{sheng2013survey}, some constrained devices may face similar effects as a Denial-of-Service (DoS) attack from huge amounts of legitimate clients' requests that may hinder the services provided. Still, current Internet Engineering Task Force (IETF) standardized communication protocols, such as Constrained Application Protocol (CoAP), failed to provide solutions and, therefore, foster IoT network availability. \\
Denial-of-Service attacks obstruct the communication between devices and prevent them for accessing network resources. According to \cite{kasinathan2013denial}, DoS attacks are important security issues that need to be addressed. DoS attacks can be executed remotely with simple commands in combination with more sophisticated tools that may allow the execution of Distributed Denial-of-Service (DDoS) attacks as well. DoS attacks against IoT systems, as exposed by \cite{roman2013features}, not only deal with traditional vectors, such as service provider resource and bandwidth exhaustion but also they can compromise data acquisition wireless communication from IoT nodes. \cite[p. 649]{suo2012security} states that DDoS are ``particularly severe" for IoT systems constituted of vulnerable nodes from a network layer standpoint. \cite{kasinathan2013denial} defines the range of DoS attacks, from the simplest one, such as jamming attacks (the interference of radio signals), to sophisticated ones, such as elaborated DDoS.
DoS and DDoS attacks not only can affect the availability of network resources or applications, but also may cause energy dissipation issues, critical for constrained devices \cite{misra2011learning}. Physical damage, as stated by \cite{roman2013features}, can also be considered a DoS threat executed by less knowledgeable attackers to cripple IoT services by ``things" destruction. \\   
\cite{misra2011learning} defines DDoS attacks as the set of concurrent DoS attacks. Therefore, the authors suggest in their work, a Service Oriented Architecture (SOA) as a DDoS prevention strategy for IoT systems. Traditional approaches to prevent DoS or DDoS attacks rely on heavy network traffic sampling, \cite{misra2011learning} proposes an optimized solution based on random sampling and sampling rate efficiency. \cite{kasinathan2013denial} lists various DoS defense techniques for WSN including \cite{raymond2008denial}, \cite{garcia2012security} and \cite{heer2011security}. Nevertheless, the authors indicate that there is not an existing defense mechanism capable of ruling out DoS risks. DoS attack detection is very difficult to accomplish, according to \cite{kasinathan2013denial}, since the symptoms of such attacks may also make some services unavailable. \cite{kasinathan2013denial}, proposes an Intrusion-Detection-System (IDS) based solution for DoS attack detection. The solution objective is to detect DoS attacks in early stages before the disruption of normal network operations for 6LoWPAN solutions. 
The authors of \cite{roman2013features} suggest the implementation of distributed architectures instead of centralized approaches. One of the main advantages portrait in the same work, is the improvement of availability properties in terms of service uptime as well as eliminating single points of failure. Suo et al. \cite{suo2012security} stress on the importance of disaster recovering procedures to be placed after large-scale or elaborated DDoS attacks.

%\section{System Constraints}

%\subsection{Device Process Power}
%\subsection{Algorithm Complexity}
%\subsection{Energy Consumption}

%\section{Challenges and Proposed Solutions}

\subsection{Privacy}

The significant growth of the IoT has showed during recent years has brought in several privacy concerns as data availability soars, sponsored by ubiquitous and pervasive properties of the IoT \cite{miorandi2012internet}\cite{stankovic2014research}, and the fact that devices at the moment do not offer all the desired warranties. \cite{sicari2015security} calls for protection of users' personal information associated to their ``movements, habits and interactions" \cite[p. 151]{sicari2015security}.  Faulty provisioning of data confidentiality and integrity could influence user privacy as malicious parties could access sensitive data without any authorization or consent\cite{medaglia2010overview}\cite{atzori2010internet}, harming as well the possibility for widespread adoption of IoT technologies \cite{atzori2010internet}\cite{miorandi2012internet}\cite{tan2010future}. \cite{roman2011securing} discusses worst-case scenarios and undesirable situations produced by ``Big Brother-like entities" \cite[p. 54]{roman2011securing} were data is collected and shared without user consent. \\
\cite{vermesan2013internet} distinguishes issues and challenges that the IoT community needs to address in order to prevent privacy violation, which includes self-aware behavior of interconnected devices, data integrity, authentication, heterogeneity tolerance, efficient encryption techniques, secure cloud computing, data ownership \& governance, as well as policy implementation and management. \cite{roman2011securing} also proposes solutions to the Iot privacy problems, the first one is to provide ``privacy by design" \cite[p. 64]{roman2011securing} which advocates for users to have the tools to dynamically control the data collected, stored and shared, user's request should be correlated and evaluated to existing policies in order to make a decision whether to grant data access or not \cite{stankovic2014research}.  \cite{weber2015internet} also includes Transparency as a privacy solution, as it allows users to know the parties that manage and utilize the data collected by a IoT device. \cite{stankovic2014research} proposes Data Management as a solution as well, composed by the implementation of differentiated policies and enforcing instruments. \cite{stankovic2014research} also discusses the necessity for data typification  as well as ownership, access extent (minimum and maximum of data to be read), anonymity and its viability. \cite{atzori2010internet} proposes the implementation of opt-out features managed by individuals wherever an untrustworthy sensor network has been implemented, also called ``right to silence of the chips" \cite[p. 26]{weber2010internet}, as well  the interaction with a ``privacy broker" \cite{lioudakis2007proxy} that acts as a proxy between the user and the network. \cite{tan2010future} indicates that technological solutions are not enough to address the current privacy issues and calls for the consideration of economical and socio-ethic aspects of the IoT environment. \cite{roman2013features} includes as well the revision of the existing privacy regulations at the private and government level as well as improving the users' awareness on how sensor-based devices collect, store and share their information. However, it is not that simple as stated by \cite{peppet2014regulating} as the re-classification and distinction between regular and Personal Identifiable Information (PII), which can be protected by law, for IoT devices is still a challenge to overcome. \cite{weber2010internet} questions whether IoT privacy regulations should be covered by governmental or by self-regulatory entities (current trend), government regulations could be only applicable locally when the nature of IoT data transcends boundaries and jurisdictions. Nevertheless, government entities, such as the European Commission and the US Federal Communications Commission (FCC) \cite{weber2015internet}, have already called for recommendations in the deployment of sensor networks as well as the collaboration within the civil society stakeholders for the creation of a privacy framework that works at different levels.   

\section{IoT Security Challenges and Some Solutions}
Different authors with different terminologies coincide on determining the architectural structure of the IoT. The perception, the network and the application layer (middleware can be placed in between the last two layers) constitute what currently the Internet of Things relay, each one of them provides significant value to the whole system. These segmentation provides modularity and helps systems to escalate more efficiently. However, it also allows malicious entities, in this context external attackers under the threat model defined, to exploit vulnerabilities intrinsic to each one of the IoT layers. Each one of the IoT components of the different layers can be run on top of separate technologies and, therefore, distinct weaknesses are found based on functionality and application. Such vulnerabilities have been exploited in a way that have compromised millions of IoT devices which have resulted in the perfect weapon to execute one of the most internet-disruptive breakdowns in recent times. Even though security researchers have expressed their concern over the weaknesses of IoT systems the intrinsic principle of energy efficiency as well as low computing power available on embedded devices are in some way antagonistic to the existing cryptography principles, that means a more challenging environment for the IoT and its community.  

The IoT, then, needs a deeper discussion that strengthens its foundations towards a secure environment. In order to contribute to this purpose, the authors of this document consider that further analysis under the security triad (confidentiality, integrity and availability) is necessary. Also, this work covers the privacy concerns that the ongoing soaring demand for IoT devices has brought along. Many IoT experts have raised their concerns on how ``Big Brother-like" entities may collect and disclose users' data without consent and how technological and governance implementations may help to relieve the existing doubts. As stated, the situation of the IoT under a security perspective is concerning and proper analysis and consequent actions are required. The need for integral standards as well as for more hardware-friendly security implementations is now of common understanding. Policy is also a priority for user protection and manufacturer regulation in order to find a more fertile ground for IoT expansion. 

\section{Conclusion}
The ongoing Internet of Things state reveals that there is still significant work to do in order to secure embedded computer devices. Even though the number of IoT devices as well as new technologies and scientific publications has soared in the last few years, the security solutions and improvements have not kept the pace. Publicly-known security breaches initiation vectors point to vulnerable and/or neglected IoT devices and the number of records stolen continue to grow. The amount of data handled by IoT devices is soaring at exponential rates, which means higher exposure of sensitive data and brings up the need to foster discussions among security researchers. Recent efforts have not been able to cover the entire security spectrum, which reveals research opportunities in different areas including smart object hardening and detection capabilities. Current issues and challenges should be taken as improvement opportunities that need to be achieved under a rigorous process that incorporates security objectives at early design stages and efficient and effective application of security standardized solutions at production stages. Final users, as well, need to understand the main objective of the device and how to fulfill their requirements under strict control and scrutiny to manage the always present risk for inter-connectivity.      

\ifCLASSOPTIONcaptionsoff
  \newpage
\fi
\bibliographystyle{ieeetr}
%\bibliography{references}

\end{document}